\documentclass[11pt,a4paper]{article}
\usepackage{jheppub,subfigure}

\title{Astrophysical neutrino flavor ratios in the presence of sterile neutrinos}
\author[a]{D. Hollander}
\affiliation[a]{Department of Physics, The Pennsylvania State University,\\
104 Davey Lab, University Park, PA, 16802-6300}
\emailAdd{daveh@phys.psu.edu}

\abstract{Astrophysical objects such as active-galactic nuclei (AGN) and gamma-ray bursts (GRBs) can be sources of high energy, astrophysical neutrinos.  The decay of charged pions produces electron and muon-flavor neutrinos from the primary decay of the pion, and from the secondary decay of the resulting charged lepton.  At low energies we expect the flavor ratio $\Phi_{\nu_e}:\Phi_{\nu_\mu}:\Phi_{\nu_\tau}$ to be $1:2:0$ at the source.  We are interested in the flavor ratios as measured on Earth after the neutrinos propagate over cosmic distance scales from the source.  If we only consider vacuum flavor transition probabilities between the three active flavors then we expect a measured flavor ratio of $1:1:1$ up to small corrections from $\theta_{13}$ and non-maximal $\theta_{23}$.  When we include mixing with two additional flavors of sterile neutrinos then we see corrections to this ratio up to $\sim 30\%$.  Furthermore, if we consider energy-loss of the charged leptons involved in the pion decay from cosmic sources then these flux ratios depend on the neutrino energy.  We examine these energy-dependent flavor ratios using a specific model for sterile neutrino mixing, and compare to expected ratios when only three neutrino flavors are considered.  At energies $E_\nu >$ 1 TeV the flavor ratios observed in experiments such as IceCube can probe the existence of sterile neutrinos. }

\begin{document}
\maketitle

\section{Introduction}
\label{sec:intro}

High-energy, charged pions are produced by p$\gamma$ or p-nucleon scattering in astrophysical objects.  The charged pions may be from gamma-ray bursts (GRBs) \cite{coolingNew,cooling,astrophys} where they are produced with a flux $\Phi_\pi \propto E_\pi^{-2}$ \cite{cooling}, and also they may result from gamma ray-jets in active galactic nuclei (AGN) \cite{astrophys,AGN1,AGN2}.  These pions produce neutrinos through the process $\pi^+ \to \nu_\mu + \mu^+ \to \nu_\mu + \nu_e + e^+ + \bar{\nu}_\mu$, so we expect a $1:2:0$ flavor ratio from the source where all ratios are given as $\Phi_{\nu_e}:\Phi_{\nu_\mu}:\Phi_{\nu_\tau}$.  Neutrino propagation is affected by flavor oscillations, and oscillation probabilities are included in the calculation of final flavor states.  Therefore measured neutrino fluxes on the Earth are flavor-mixing model dependent.  When we only consider three flavors of neutrinos, we expect a flavor ratio $1:1:1$ as measured on the Earth \cite{flavor_ratio1,flavor_ratio2}, up to corrections from a non-zero $\theta_{13}$, and non-maximal $\theta_{23}$ \cite{flavor_ratio4}.

We must consider cooling effects from astrophysical sources on the charged leptons and pions which produce these neutrinos.  Energy losses factor in from astrophysical sources such as losses due to synchrotron radiation \cite{coolingNew,cooling} or inverse-Compton emission \cite{cooling}, as in GRBs, or adiabatic cooling as in AGN \cite{cooling,AGN2}.  These losses give the measured ratios a neutrino-energy dependence.  Different models of energy-losses lead to different flavor ratio profiles; measurements of these neutrino flavor ratios over several decades of neutrino energies may allow one to determine the source of the neutrinos depending on the profile.

In this paper we will consider the expected neutrino flavor ratios when couplings to two additional flavors of sterile neutrinos are included.  Corrections to flavor ratios when one additional flavor of sterile neutrinos is included have been considered before in \cite{flavor_ratio2,flavor_ratio3}.   In section \ref{sec:cooling_models} we will discuss different models for the cooling effects of the charged pions and leptons.  In section \ref{sec:prob} we will describe the formalism of neutrino flavor-oscillations, and briefly describe the model of sterile neutrino mixing used here.  Finally, we will compare the predictions from a five-flavor case to the three-flavor case of the Standard Model in section \ref{sec:earth_flux}.

\section{Models for charged pion/lepton energy loss: source effects}
\label{sec:cooling_models}

We now want to consider different models of charged pion production as well as energy-loss for charged pions and charged leptons prior to their decay.

We consider an astrophysical source which produces charged pions with a flux $\Phi_\pi \propto E_\pi^{-k}$.  Energy losses are also assumed to follow the power law $\dot{E}_x = dE_x/dt \propto -E_x^n$.  If we consider pions produced by the process $p+\gamma \to n + \pi$ in a GRB then we have a flux of pions which behaves as $\Phi_\pi \propto E_\pi^{-2}$ \cite{coolingNew,cooling}.  The presence of strong intergalactic electromagnetic fields lead to radiative energy losses of the pion and subsequent charged leptons prior to decay.  Synchrotron energy loss, and likewise inverse-Compton emission, correspond to $n=2$ \cite{cooling}.  If we consider adiabatic energy loss as a result of expansions of the $\pi^+/\mu^+$ plasma then this corresponds to $n=1$ \cite{cooling}.  In the process $p+\gamma \to n + \pi$, although the neutron can decay to produce $\bar{\nu}_e$, the fraction of the proton's energy that this neutrino carries is much smaller than the fraction carried by the neutrinos resulting from pion decay.  We will therefore ignore the contribution to fluxes from the resulting neutron.

The energy dependent fluxes of $\nu_\mu, \nu_e$ and $\bar{\nu}_\mu$ at the source are calculated for example in \cite{cooling}, and the results are given here.  For the production of $\nu_\mu$ from the primary decay of $\pi^+$ we have
\begin{equation}
\label{eqn:primary}
\Phi^s_{\nu_\mu}(E_\nu)=s_n(-s_n)^{(1-k)/n}e^{-s_n}\Gamma\left( \frac{k-1}{n},0,-s_n \right)
\end{equation}
For the production of $\bar{\nu}_\mu$ and $\nu_e$ from the secondary decay of the $\mu^+$ we have a source flux
\begin{equation}
\begin{aligned}
\label{eqn:secondary}
\Phi^s_{\bar{\nu}_\mu,\nu_e}(E_\nu)=\frac{1}{q^{-n}-1}(-s_n)^{(1-k+n)/n} &\left\{ e^{-s_n}\Gamma\left( \frac{k-1}{n},0,-s_n \right) \right. \\
&\left. - q^{1-k}e^{-q^n s_n}\Gamma\left( \frac{k-1}{n},0,-q^n s_n \right) \right\}
\end{aligned}
\end{equation}
where $s_n=\frac{1}{n} \left(\frac{E_{\pi,\textrm{cool}}}{4E_\nu} \right)^n$, $q=\frac{4 E_{\mu,\textrm{cool}}}{3E_{\pi,\textrm{cool}}}$, and $\Gamma\left( \frac{k-1}{n},0,-s_n \right)$ is a lower-incomplete gamma function.  $E_{x,\textrm{cool}}$ is the energy at which the time, $\tau_{x,\textrm{cool}}$ to achieve significant energy loss (cooling) due to synchrotron radiation or adiabatic energy loss is the same as the decay time, $\tau_{x,\textrm{decay}}$ \cite{cooling}.  Energy loss from astrophysical sources changes quickly with energy, such that $\tau_{x,\textrm{cool}}\propto E_x / \dot{E}_x$, and so $\tau_{x,\textrm{cool}}/ \tau_{x,\textrm{decay}} \propto E_x^{-n}$, and also $(E_{\pi, \textrm{cool}} / E_{\mu, \textrm{cool}} ) \sim 10^{2/n}$ \cite{cooling}.

In the next section we will use the energy-dependent neutrino flux at the source to calculate fluxes on the Earth of the three active neutrinos when couplings to two sterile neutrinos in the 3+2 MM are considered.

\section{Neutrinos oscillations/propagation effects and the 3+2 minimal model}
\label{sec:prob}

Neutrino flavor-states are linear combinations of the different mass eigenstates.  Given $N$ flavors of neutrinos, there are $N$ neutrino mass eigenstates, and mixing is determined by a unitary $N \times N$ matrix, $U$.  Generically this mixing is given by
\[
| \nu_\alpha \rangle = \sum_i U_{\alpha i} |\nu_i \rangle
\]
where $U$ is parameterized by several mixing angles and phases; given $N$ neutrino flavors there are $N(N-1)/2$ mixing angles, and up to$N(N+1)/2$ phases.  In the case of the $3\nu$ of the Standard Model the matrix $U$ is given by the PMNS matrix, parameterized by current global fits given in \cite{global_fits}.

We can compute oscillation probabilities for the process $\nu_\alpha \to \nu_\beta$; probabilities depend on the neutrino energy as well as the length of propagation, $L$, and are given by
\begin{equation}
\begin{aligned}
\label{eqn:prob}
P_{\alpha\beta}(L/E)=\delta_{\alpha\beta} - 4\sum_{i>j} \textrm{Re}&(U^*_{\alpha i} U_{\beta i} U^*_{\beta j} U_{\alpha j}) \sin^2\left(\frac{\Delta m_{ij}^2L}{4E}\right) \\
+ 2&\sum_{i>j} \textrm{Im}(U^*_{\alpha i} U_{\beta i} U^*_{\beta j} U_{\alpha j}) \sin\left(\frac{\Delta m_{ij}^2L}{2E}\right)
\end{aligned}
\end{equation}
We are neglecting potential terms which arise from neutral-current and charged-current weak interactions with matter, and only considering vacuum oscillations.  Matter potentials are proportional to the density of electrons or nucleons in the propagating medium, and these densities are very low in most astrophysical cases which makes matter potentials negligible.  Some astrophysical objects, such as the fireballs from GRBs are sufficiently larger that only very light neutrinos are affected, therefore it is a valid choice to exclude any matter interactions.  For example, in a GRB fireball the electron density is $\sim 10^{10} - 10^{12}\textrm{cm}^{-3}$, and so the MSW resonance would only significantly affect neutrinos with mass differences less than $\Delta m^2 \sim 10^{-12}\textrm{ eV}^2$ \cite{fireball}.

Since we are considering neutrinos from astrophysical sources we have a very large propagation length, and therefore we are in the limit $x=\Delta m_{ij}^2 L/4E \gg 1$.  In this limit oscillations are very rapid, and oscillation terms in the probability take on their average values, $\sin^2(x)\to \frac{1}{2}$ and $\sin(2x)\to 0$.  Therefore the probabilities in the flux calculations become
\begin{equation}
\label{eqn:prob_limit}
P_{\alpha \beta}=\langle P_{\alpha\beta}(L/E) \rangle= \delta_{\alpha\beta}-2\sum_{i>j} \textrm{Re}(U^*_{\alpha i} U_{\beta i} U^*_{\beta j} U_{\alpha j}) = \sum_{i} |U_{\alpha i}|^2 |U_{\beta i}|^2 
\end{equation}

\begin{table}
\centering
\begin{tabular}[ht!!]{l|l|l|c|c|c|c|c} \hline
3+2 MM & $|\Delta m_{41}^2|(\textrm{ eV}^2)$ & $|\Delta m_{51}^2|(\textrm{ eV}^2)$ & $|U_{e4}|$ & $|U_{e5}|$ & $|U_{\mu 4}|$ & $|U_{\mu 5}|$ & $\phi_{45}$ \\ \hline
NH & 0.47 & 0.87 & 0.149 & 0.127 & 0.112 & 0.127 & 1.8$\pi$ \\
IH & 0.9 & 1.61 & 0.139 & 0.122 & 0.138 & 0.107 & 1.4$\pi$ \\ \hline
\end{tabular}
\caption{Results of the fits for the 3+2 MM for both mass hierarchies. \label{tab:fit}}
\end{table}
The model for sterile neutrino mixing used here involves two sterile neutrinos, with two additional sterile mass eigenstates where $m_{4,5} \sim \mathcal{O}(\textrm{eV}^2)$ \cite{model}.  This model is a minimal extension of the Standard Model because it involves adding only two Standard Model gauge singlet Weyl fields; each field corresponds to a unique sterile flavor.  The additional terms in the Lagrangian are given by
\[
\mathcal{L}_{BSM}= -\bar{l}^\alpha_L Y^{\alpha j}\Phi \nu^j_R - \frac{1}{2} \bar{\nu}_R^{ic}M_R^{ij}\nu_R^j + h.c.
\]
where index summation is implied, and $c$ indicates charge conjugation.  $Y$ is the $3\times 2$ Yukawa matrix, $\Phi$ is a scalar, SSB field, and $M_R$ is a diagonal, $2\times 2$ mass matrix.  A basis is chosen for this model such that the first mass eigenstate becomes massless, and $\Delta m_{32}^2 = \Delta m_{31}^2$.

The details of the parameterization and fits for this model can be found in the original paper \cite{model}.  To summarize, the 3+2 MM is parameterized by four mixing angles (the three angles of the Standard Model and one additional angle which mixes the two sterile mass eigenstates), three phases which includes a relative phase between the sterile mass eigenstates, and four non-zero mass eigenstates.  The results of the parameterization are summarized in table \ref{tab:fit}.  The authors in \cite{model} found that the normal hierarchy provides a better fit to neutrino flux data which includes anomalies, and therefore we will work in the normal hierarchy (NH) for the remainder of this paper.

\section{Active neutrino flavor ratios on the Earth}
\label{sec:earth_flux}

\begin{figure}[ht!!!]
\centering
\subfigure[\label{fig:flux_ratios}]{\includegraphics[width=0.45\textwidth]{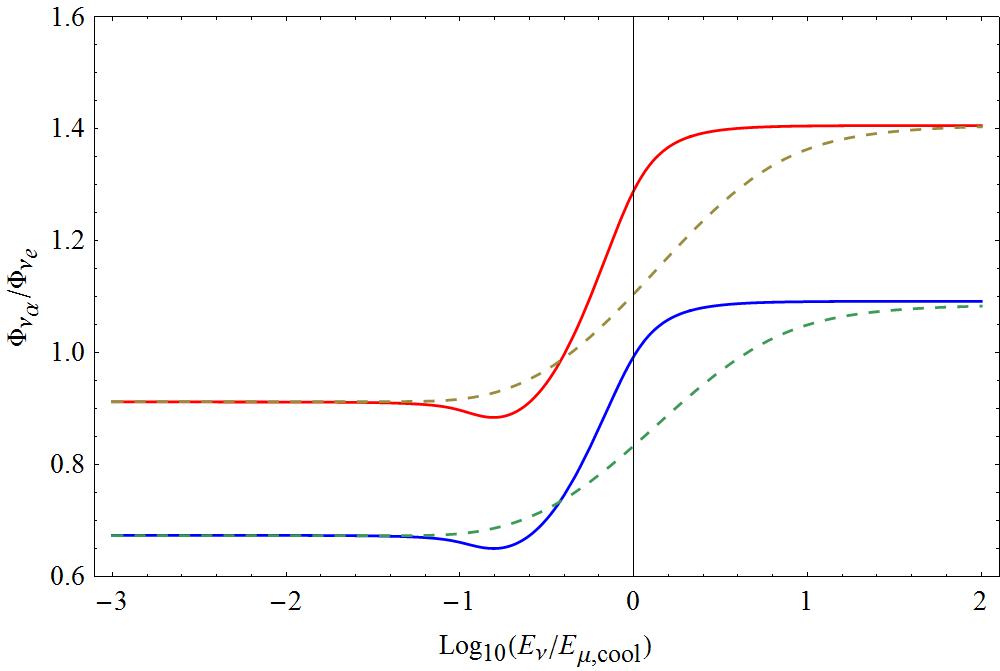}}
\subfigure[\label{fig:theory_ratios}]{\includegraphics[width=0.45\textwidth]{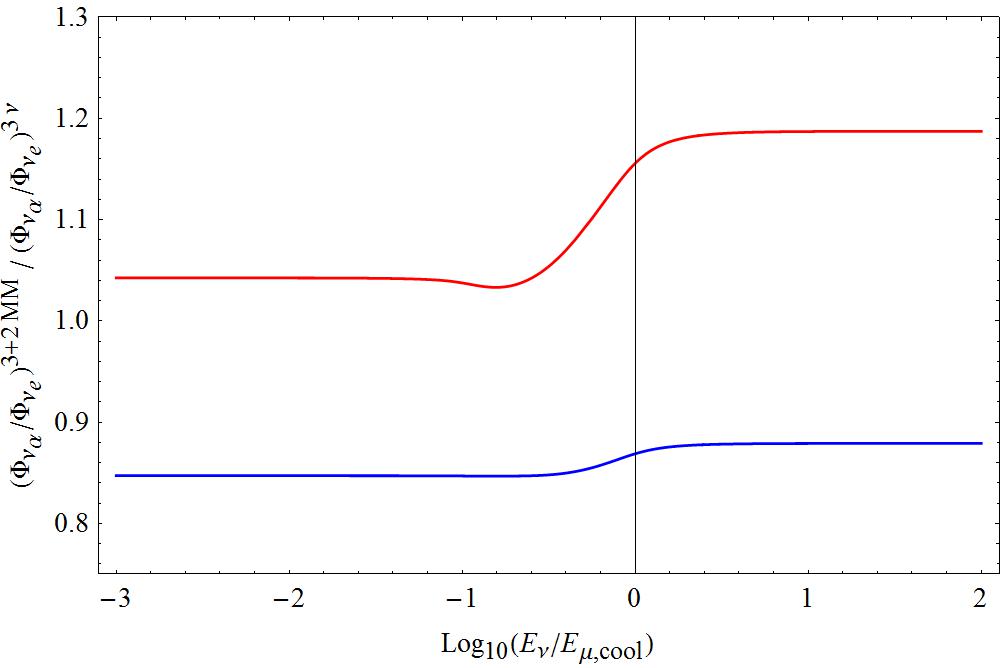}}
\caption{(a) A plot of $\Phi_{\nu_\alpha}/ \Phi_{\nu_e}$.  The solid curves consider fluxes where the $\pi^+$ and $\mu^+$ energy losses are due to synchrotron radiation or reverse-Compton emission (n=2); the solid red curve (upper) is $\alpha=\mu$, while the solid blue curve (lower) is $\alpha=\tau$.  The dashed curves consider fluxes where the $\pi^+$ and $\mu^+$ energy losses are adiabatic (n=1); the upper dashed curve is $\alpha=\mu$, while the lower dashed curve is $\alpha=\tau$.  Flux from $\pi^+$ and $\pi^-$ decay are considered here. (b) The ratio $(\Phi_{\nu_\alpha}/ \Phi_{\nu_e})^{3+2 \textrm{ MM}} / (\Phi_{\nu_\alpha}/ \Phi_{\nu_e})^{3\nu}$.  The red (upper) is $\alpha=\mu$, and the blue (lower) curve is $\alpha=\tau$.  The ratios are calculated for n=2 only.}
\end{figure}

As the neutrinos are allowed to propagate from the source to the detector on Earth there will be transitions between flavor states.  The final states measured on the Earth are given by
\[
\Phi^d_{\nu_\alpha}(E_\nu)=\sum_\beta P_{\alpha\beta}\Phi^s_{\nu_\beta}(E_\nu)
\]
where transition probabilities used here are given by the limiting case in eqn. \ref{eqn:prob_limit}.  We will be examining the case where the pions at the source are produced in a gamma-ray burst, and the flux goes as $\Phi_\pi \propto E_\pi^{-2}$ (k=2).  Plots of the flavor ratios for different models of pionic, and leptonic energy losses can be seen in Fig. \ref{fig:flux_ratios}, and the comparison to the three flavor case of the Standard Model can be seen in Fig. \ref{fig:theory_ratios}.

\begin{figure}[ht!!!]
\centering
\subfigure[\label{fig:flux_ratios_GRB}]{\includegraphics[width=0.45\textwidth]{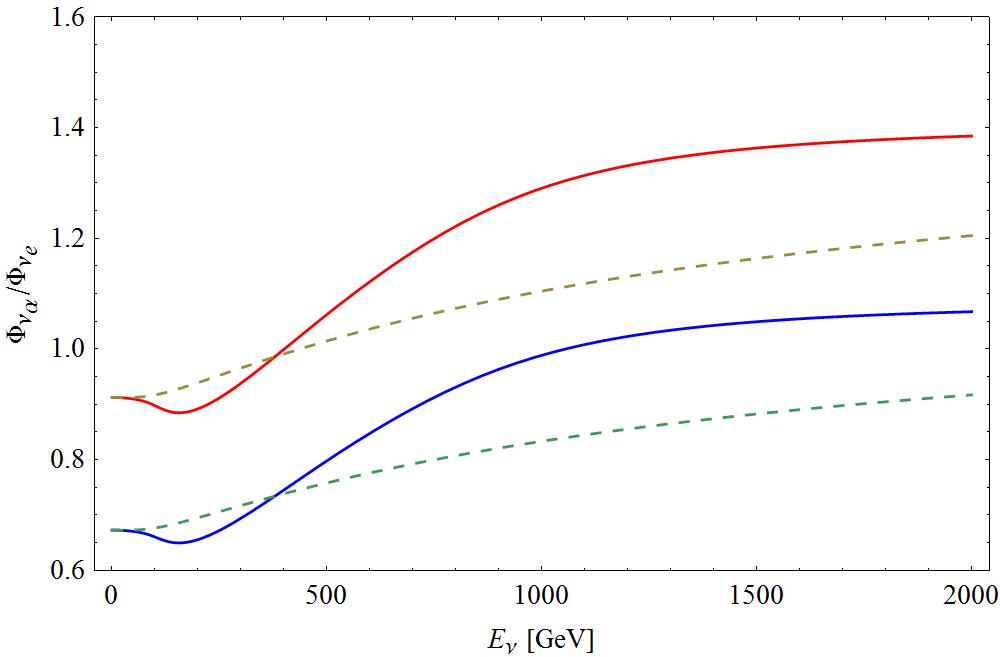}}
\subfigure[\label{fig:theory_ratios_GRB}]{\includegraphics[width=0.45\textwidth]{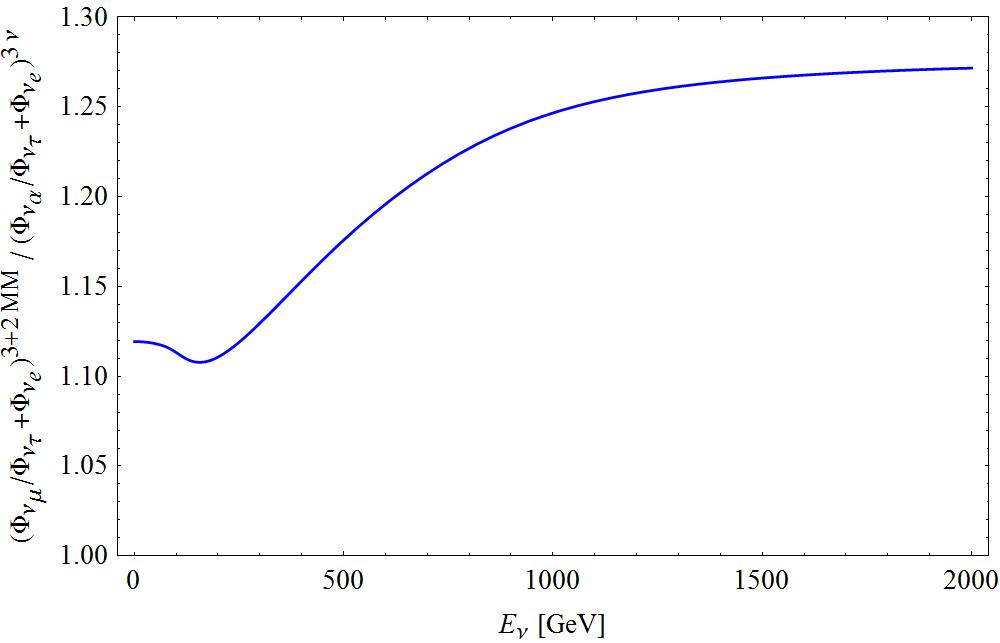}}
\caption{(a) A plot of the same quantity as in Fig. \ref{fig:flux_ratios}, where $E_{\mu,\textrm{cool}}=$ 1 TeV.  This muon cooling energy approximately corresponds to a model of gamma-ray bursts associated with the collapse of a massive star \cite{cooling}.  (b) Comparing the ratio of tracks ($\nu_\mu$) to cascades ($\nu_\tau + \nu_e$) between the 3+2 MM and the prediction with only three flavors where $E_{\mu,\textrm{cool}}=$ 1 TeV.}
\end{figure}

We can see the flavor transition in the neighborhood around $E_\nu = E_{\mu, \textrm{cool}}$ where the flux ratios transition from $\sim 1:0.9:0.7$ to $\sim 1:1.4:1.1$.  Determining $E_{\mu, \textrm{cool}}$ is model-dependent.  An interesting feature can be seen in Fig. \ref{fig:theory_ratios}, where the ratio $\Phi_{\nu_\mu}/ \Phi_{\nu_e}$ predicted by the 3+2 MM is larger than the ratio predicted by transitions between the three flavors of the Standard Model over all decades of neutrino energies, but especially at high energies $\log_{10} (E_\nu / E_{\mu,\textrm{cool}}) \gtrsim 0.5$.  This implies larger probabilities of $P(\nu_\alpha \to \nu_\mu)$ via some process such as $\nu_\alpha \to \nu_s \to \nu_\mu$ involving one or more transitions from active to sterile back to active flavors of neutrinos.  This predicts that we would expect the measured $\nu_\mu$ flux to be larger at ultra-high energies, and the ratio $\Phi_{\nu_\mu}/\Phi_{\nu_e}$ is significantly higher with two sterile neutrinos than the same quantity for only three flavors.  This would provide a clear test for the presence of sterile neutrinos in the 3+2 MM scheme.  Flavor ratios as a function of neutrino energy when $E_{\mu,\textrm{cool}}=$ 1 TeV can be seen in Fig. \ref{fig:flux_ratios_GRB};  $E_{\mu,\textrm{cool}}=$ 1 TeV approximately corresponds to a model of gamma-ray bursts associated with the collapse of a massive star where cooling is a result of inverse-Compton emission \cite{cooling}.

For energies $E_\nu \lesssim$ 1 TeV it is not possible to separate electron neutrinos from tau neutrinos.  In detectors such as IceCube \cite{icecube} muon neutrinos produce tracks whereas electron and tau neutrinos contribute to cascades at these low energies.  At high energies it may be possible to differentiate between electron and tau flavors by observing lollipops or $\nu_\tau$ charged-current double-bangs \cite{doublebang}.  The comparison of the tracks to cascades ratio between the 3+2 MM and three flavors can be seen in Fig. \ref{fig:theory_ratios_GRB}, where the energy region prohibits the differentiation between the tau and electron flavors. 

\section{Conclusions}
\label{sec:conclusion}

Examining neutrino flavor ratios as a function of the neutrino energy can potentially allow one to determine the source of the neutrinos.  Shapes corresponding to $k=2$ and $n=2$ as in the solid curves of Fig. \ref{fig:flux_ratios} may indicate a GRB source, whereas shapes corresponding to $k=2$ and $n=1$ as in the dashed curves may indicate an AGN source.  Calculating flux ratios also has the potential for determining whether there are flavors of sterile neutrinos.

The flux predictions in this paper also provide a method for determining the existence of sterile neutrinos, specifically whether we have a minimal extension of the Standard Model involving two sterile flavors.  If GRBs are associated with the collapses of massive stars then $E_{\mu,\textrm{cool}} <$ 1 TeV due to inverse-Compton emission\cite{cooling}, and therefore experiments such as IceCube \cite{icecube} which can measure high-energy neutrinos over several energy decades are good candidates for measuring these flavor ratios.

Verifying these flavor ratio profiles requires statistics from many neutrino events.  Nearby gamma-ray bursts which provide more than 1 - 2 neutrinos are infrequent ($\sim 1/\textrm{century}$), and GRBs typically have durations of 0.1 - 100 seconds \cite{astrophys} so the number of neutrino events from a single GRB will be small; it is necessary to take measurements from many GRB sources.  Although AGN emit high energy radiation over a period lasting weeks \cite{astrophys} the number of neutrino events expected from a single AGN is low.  In both GRB and AGN cases we must wait a long time in order to measure enough events to extract flavor ratios from background.

\section{Acknowledgments}

We would like to thank P. M\'{e}sz\'{a}ros for his suggestion to investigate this problem, as well as his advice on what would be interesting to examine from an astrophysical standpoint.  We would also like to thank I. Mocioiu for her guidance in this research.

\end{document}